\documentclass[prb,twocolumn,showpacs,amsfonts,amssymb,amsmath,superscriptaddress]{revtex4} 

	\usepackage{bm}
	\usepackage{braket}
	\usepackage{dsfont}

 	\usepackage{graphicx}
	\usepackage[caption=false]{subfig}

	\newcommand{\vect}[1]{\boldsymbol{#1}}		
	\newcommand{\op}[1]{\hat{\boldsymbol{#1}}}	
	
			\graphicspath{ {FinalDiagrams/} }

\begin{document}

\title{Controlled formation of an isolated miniband in bilayer graphene on an almost commensurate $\sqrt{3} \times \sqrt{3}$ substrate}
\author{D.~J.~Leech}
\email{D.J.Leech@bath.ac.uk}
\affiliation{Department of Physics, University of Bath, Claverton Down, Bath BA2 7AY, United Kingdom}
\author{M.~Mucha-Kruczy\'{n}ski}
\affiliation{Department of Physics, University of Bath, Claverton Down, Bath BA2 7AY, United Kingdom}
\affiliation{Centre for Nanoscience and Nanotechnology, University of Bath, Claverton Down, Bath BA2 7AY, United Kingdom}

\begin{abstract}
We investigate theoretically the interplay between the effects of a perpendicular electric field and incommensurability at the interface on the electronic properties of a heterostructure of bilayer graphene and a semiconducting substrate with a unit cell almost three times larger then that of graphene. It is known that the former introduces an asymmetry in the distribution of the electronic wave function between the layers and opens a band gap in the electronic spectrum. The latter generates a long wavelength periodic moir\'{e} perturbation of graphene electrons which couples states in inequivalent graphene Brillouin zone corners and leads to the formation of minibands. We show that, depending on the details of the moir\'{e} perturbation, the miniband structure can be tuned from that with a single band gap at the neutrality point and over-lapping minibands on the conduction/valence band side to a situation where a single narrow miniband is separated by gaps from the rest of the spectrum.
\end{abstract}
\maketitle

\section{Introduction}

Since the seminal paper by Esaki and Tsu, \cite{esaki_ibm_1970} the idea of tailoring the electronic properties of materials by forming superlattices has had a huge impact on semiconductor physics. \cite{tsu_suptonan_2011} More recently, various works studied the possibility of modulating the electronic properties of graphene by applying a lateral periodic potential. \cite{park_prl_2008_extpot, barbier_prb_2010_extradirac, killi_prl_2011_blgsuperlat, burset_prb_2011_mlgsuperlat, tan_nanlett_2011_blgsuperlat} Experimentally, a one-dimensional artificial graphene superlattice has been fabricated using electrostatic gates. \cite{dubey_nanolett_2013} However, a two-dimensional superlattice can also be produced by placing graphene on a hexagonal substrate/surface facet such as hexagonal boron nitride (hBN), \cite{xue_natmat_2011_stmgrahbn} Ir(111) \cite{pletikosic_prl_2009} or Ru(0001). \cite{martoccia_prl_2008} In this case, the superlattice, known as a moir\'{e} pattern, arises due to the mismatch between the graphene and the substrate lattice constants and misalignment of the crystalline directions of the two materials. For graphene on hBN, a heterostructure which attracted considerable attention because of, for example, the first observation of the fractal spectrum of magnetic minibands known as Hofstadter’s butterfly \cite{ponomarenko_nat_2013_grasuperlat, hunt_sci_2013_butterfly, dean_nar_2013_butterfly} and detection of topological valley currents, \cite{gorvachev_sci_2014_currentsuperlat} weak coupling between the two crystals and close match of the reciprocal lattice vectors in the two materials allow a continuum model description of the perturbation using only harmonic functions of the six smallest reciprocal lattice vectors of the superlattice. \cite{wallbank_prb_2013} The Bragg scattering of graphene electrons by reciprocal lattice vectors of hBN leads to the formation of minibands due to an effective coupling of states in the vicinity of the same graphene Brillouin zone (BZ) corner. \cite{wallbank_prb_2013, wallbank_adp_2015}

\begin{figure}
\subfloat{  \includegraphics[clip,width=0.90\columnwidth]{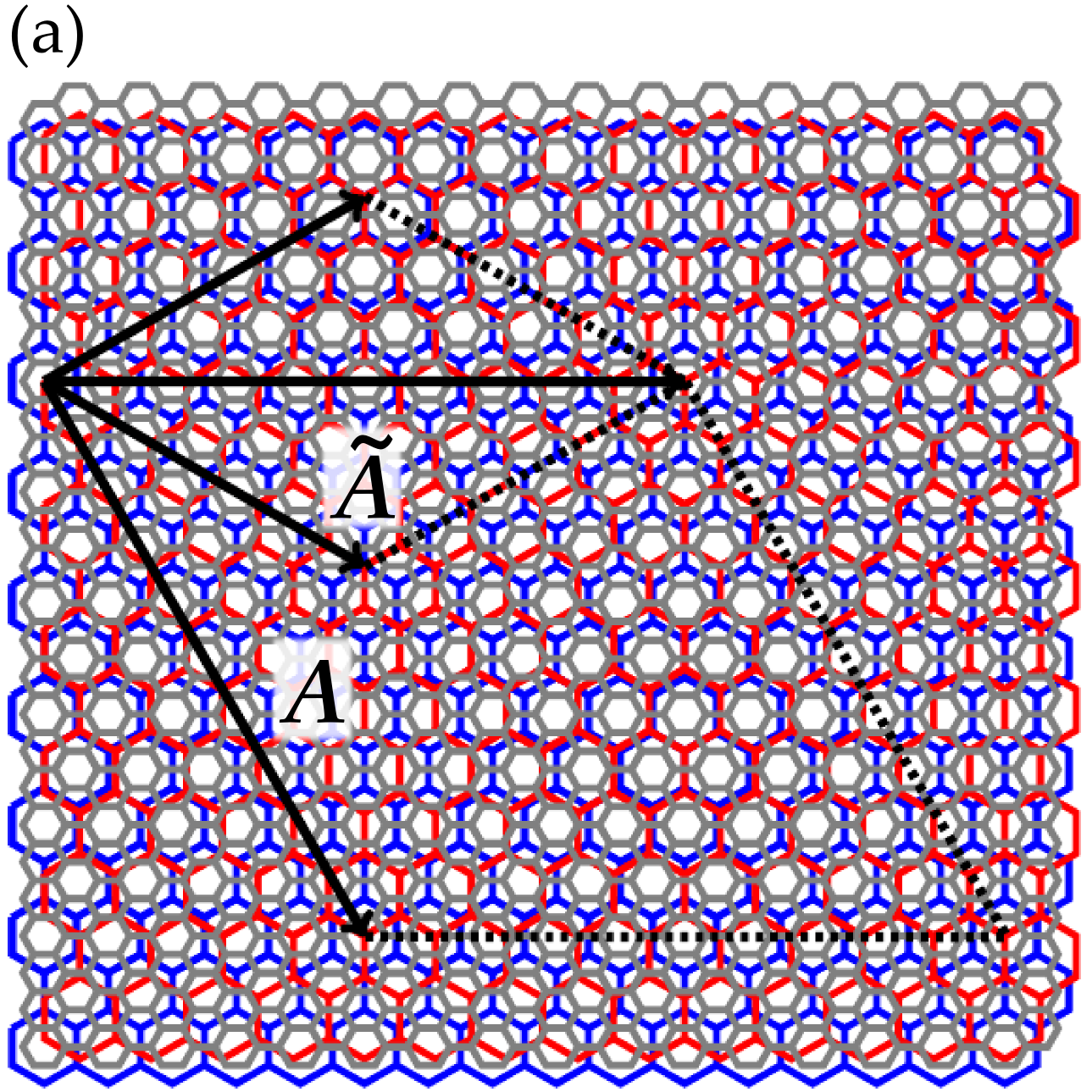} } \\
\subfloat{  \includegraphics[clip,width=0.90\columnwidth]{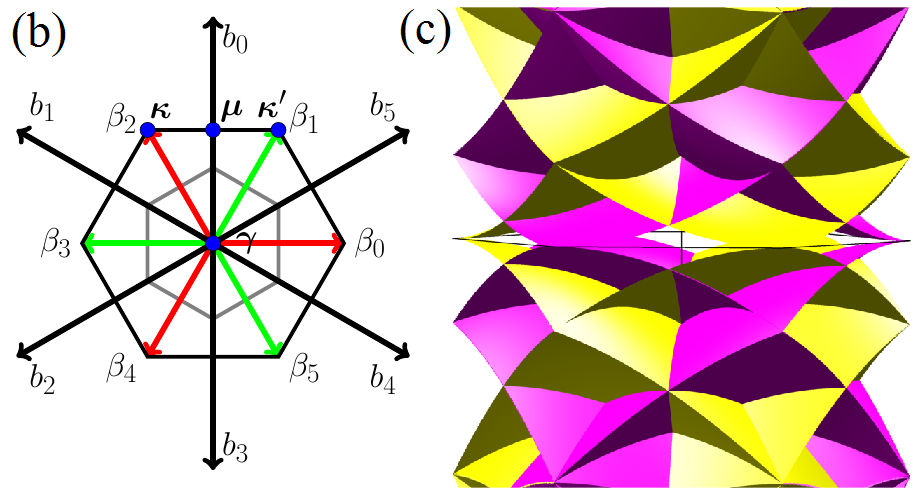} } \\
\caption{(a) Schematic of the moir\'{e} superlattice of graphene (grey) on an incommensurate $\sqrt{3} \times \sqrt{3}$ substrate (red). Also shown is the Kekul\'{e} lattice of graphene (in blue). For clarity, we choose a large $\delta$, set $\theta = 0$ and do not show the top graphene layer of BLG. While the mismatch between the red and blue lattices results in a superlattice periodicity set by $\vect{A}$, a shorter periodicity $\tilde{\vect{A}}$ of the local atomic arrangement surrounding a carbon atom always exists (see the text and Appendix for details). (b) The two periodicities set by $\vect{A}$ and $\tilde{\vect{A}}$ result in two sets of basic reciprocal vectors, $\vect{\check{\beta}}_{m}$ and $\vect{\check{b}}_{m}$, $m = 0, 1, ..., 5$, respectively, and two superlattice Brillouin zones that can be used to describe the miniband spectrum. For the larger BZ set by the reciprocal vectors $\vect{\check{b}}_{m}$, we introduce the high-symmetry points $\gamma$, $\mu$ and $\kappa$. (c) Spectrum of bilayer graphene folded onto the sBZ set by vectors $\vect{\check{b}}_{m}$. States from the $\vect{K}_{+}$ ($\vect{K}_{-}$) valley are shown in magenta (yellow).}
\label{fig:figure1}
\end{figure}

A contrasting case of coupling electronic states in the vicinity of the inequivalent graphene BZ corners can be achieved by engineering a $\sqrt{3}\times\sqrt{3}$ superlattice, \cite{farjam_prb_2009, cheianov_ssc_2009_kekdist, giovannetti_prb_2015, ren_prb_2015, venderbos_prb_2016} also called the Kekul\'{e} lattice of graphene, for example by an appropriate choice of the substrate. For such superlattice, a band gap is opened at the Dirac point. \cite{farjam_prb_2009, cheianov_ssc_2009_kekdist} It has also been suggested that for specific superlattice parameters a single-valley quadratic band crossing appears in the spectra. \cite{giovannetti_prb_2015, ren_prb_2015, venderbos_prb_2016}. If the substrate is not ideally commensurate, a long wavelength moir\'{e} pattern similar to that for graphene on hBN and shown schematically in Fig.~\ref{fig:figure1}(a) appears and the intervalley coupling oscillates in space with the moir\'{e} period. As shown recently, in such case the electron states at the Dirac point remain unaffected but typically gaps are open between the first and second miniband on the conduction/valence side. \cite{wallbank_prb_2013_root3}

In this article, we investigate theoretically the electronic properties of a heterostructure of bilayer graphene (BLG) and a semiconducting almost commensurate $\sqrt{3} \times\sqrt{3}$ substrate. We use the form of the moir\'{e} perturbation derived previously for monolayer graphene to study the miniband spectrum, in particular in the presence of an external electric field perpendicular to the graphene layers. This electric field modifies the BLG electronic spectrum \cite{mccann_prb_2006_extpot, castro_prl_2007_blgextpot, zhang_nature_2009_tunableBLG} which is folded into minibands by the moir\'{e} perturbation and redistributes the electronic wave function between the two layers, influencing the impact of the superlattice on the electronic spectrum. We show that, for a large range of moir\'{e} perturbation parameters, the miniband spectrum can be tuned via the external potential from a system with a single band gap to one with a narrow miniband separated by a band gap on each side from the rest of the spectrum. Such an effect was not predicted for BLG on hBN, for which the behaviours of the first and second miniband edges were essentially unaffected by the external electric field. \cite{mucha-kruczynski_prb_2013}

\section{Electronic Hamiltonian}

We consider bilayer graphene \cite{mccann_prl_2006}, two coupled honeycomb layers of carbon atoms in an $AB$ (Bernal) stacking, placed a substrate with a lattice constant $a_{s} = \sqrt{3} (1+\delta) a$, $|\delta| \ll 1$, where $a=2.46\AA$ is the lattice constant of graphene, with an angle $\theta$ between the crystalline directions of the two materials. The BLG unit cell contains four atoms $A1$, $B1$, $A2$ and $B2$ where $A$ and $B$ denote the two sublattices within a single layer and the numbers $1$ and $2$ indicate the bottom and top layer, respectively. To describe the electronic properties of BLG, we use the four-band model for the $\pi$-electrons, applicable in the vicinity of the BZ corner (often referred to as valley) $\vect{K}_{\xi}=\xi(\tfrac{4\pi}{3a},0)$, where $\xi=\pm 1$ distinguishes between the two inequivalent BZ corners. Because of the exponential decay of the $2p_{z}$ orbital wave function with increasing distance, \cite{eisberg_quanatom_1985} we assume that the influence of the substrate on BLG is effectively limited to the bottom graphene layer which is closer to the substrate. We follow the symmetry-based analysis performed for monolayer on an incommensurate $\sqrt{3}\times\sqrt{3}$ substrate \cite{wallbank_prb_2013_root3} and use the six shortest reciprocal lattice vectors of the moir\'{e} given by the difference between the reciprocal lattice vectors of the substrate and the Kekul\'{e} lattice of graphene, $\{\vect{\beta}_{m}\} = \hat{R}_{\tfrac{2 \pi m}{6}} \big[1 - \hat{R}_{\theta} / (1 + \delta) \big] \left( \tfrac{4\pi}{3a}, 0 \right)$, shown in Fig. \ref{fig:figure1}(b), to write the Hamiltonian
\begin{align}\begin{split} \label{HamilAndPert}
\op{H} &= \begin{pmatrix}
\op{H}_{\mathrm{MLG}} + \frac{u}{2} + \delta \! \op{H} & \op{T} \\
\op{T}^{\dagger} & \op{H}_{\mathrm{MLG}} - \frac{u}{2} \\
\end{pmatrix}, \\
\op{H}_{\mathrm{MLG}} &= v \vect{\sigma} \cdot \vect{p}, \\
\op{T} &= \frac{1}{2} \gamma_{1} (\tau_{z}\sigma_{x} - i \sigma_{y}), \\
\delta \! \op{H} &= U_{E'} v \beta F(\vect{\check{\beta}}) \sigma_{z} + U_{G} v [\nabla F (\vect{\check{\beta}})] \cdot[ \vect{\sigma}\times \textbf{l}_{z}] \\
&+ U_{G'} v [\nabla F (\vect{\check{\beta}})]\cdot\vect{\sigma}, \\
F(\vect{\check{\beta}}) &= f_{1}(\vect{\check{\beta}}) \tau_{x} + f_{2}(\vect{\check{\beta}}) \tau_{y}, \\
f_{1}(\vect{\check{\beta}}) &= \!\!\!\!\! \sum_{m=0, ..., 5} \!\!\!\!\!\! e^{i \vect{\beta}_{m} \cdot \vect{r}}, \\
f_{2}(\vect{\check{\beta}}) &= i \!\!\!\!\! \sum_{m=0, ..., 5} \!\!\!\!\! (-1)^{m} e^{i \vect{\beta}_{m} \cdot \vect{r}}.  
\end{split}\end{align}

Above, we have written the Hamiltonian $\op{H}$ in the basis of Bloch states $\{ \psi_{A1}^{K_{+}}, \psi_{B1}^{K_{+}}, \psi_{B1}^{K_{-}}, -\psi_{A1}^{K_{-}}, \psi_{A2}^{K_{+}}, \psi_{B2}^{K_{+}}, \psi_{B2}^{K_{-}}, -\psi_{A2}^{K_{-}} \}^{T}$. We have also set $\hbar=1$ and introduced two sets of Pauli matrices, $\sigma_{i}$, $\bm{\sigma}=(\sigma_{x},\sigma_{y})$ and $\tau_{i}$, acting in the sublattice and valley space, respectively, as well as their direct products $\tau_{i}\sigma_{j}\equiv\tau_{i}\otimes\sigma_{j}$. The diagonal intralayer blocks $\op{H}_{\mathrm{MLG}}$ with Fermi velocity \cite{jiang_prl_2007} $v \simeq 10^{6}$ms$^{-1}$ and electron momentum $\vect{p}$ measured from the centre of the valley, correspond to the Dirac-like Hamiltonian for electrons in monolayer graphene. The off-diagonal block $\op{T}$, with $\gamma_{1} \simeq 0.38$eV, \cite{kuzmenko_prb_2009} describes the coupling between the layers and $u$ denotes the interlayer asymmetry due to an external perpendicular electric field. The moir\'{e} perturbation is captured by the term $\delta \! \op{H}$ which appears in the top left block of the Hamiltonian $\op{H}$, corresponding to the bottom graphene layer. 

In the absence of the perturbation, $\delta \! \op{H}=0$, and for $u=0$, the Hamiltonian in Eq.~(1) results in four bands, two of which are degenerate at the points $\vect{K}_{+}$ and $\vect{K}_{-}$ at the energy corresponding to the position of the chemical potential in the charge neutral structure (the neutrality point) which we use as the zero of our energy scale. The other two bands are split by $\pm\gamma_{1}$ away from the neutrality point. \cite{mccann_prl_2006} The external perpendicular electric field breaks the layer symmetry and induces an on-site energy difference between the two graphene layers described by $u$. This leads to opening of a band gap $E_{g}\approx u$ (if $u < \gamma_{1}$) in the electronic spectrum. \cite{mccann_prb_2006_extpot}

To arrive with a form of the moir\'{e} perturbation as in Eq.~\eqref{HamilAndPert}, we assumed that the hexagonal monoatomic layer directly under graphene has inversion symmetry (interaction with atoms deeper in the bulk of the substrate are neglected because of the rapid decay of the $2p_z$ wave functions). We also do not consider any intravalley terms as these were studied before. \cite{mucha-kruczynski_prb_2013, comment} The relative strength of the perturbation, measured here in the units of $\sqrt{3}v\beta$, $\beta=|\vect{\beta}_{i}|$, is set by three dimensionless parameters $U_{E'}$, $U_{G}$, $U_{G'}$. Their exact values depend on the substrate as well as the misalignment angle $\theta$ and are difficult to determine due to the van der Waals nature of the interaction between the two constituent materials (for example, multiple models with different outcomes have been suggested to describe the moir\'{e} perturbation in the graphene/hBN heterostructure \cite{wallbank_adp_2015, mucha-kruczynski_prb_2016}). However, we assume that the perturbation parameters are small, such that $|U_{i}|\ll 1$. Finally, we note that for $\theta=0$ the reflection axes of the graphene and substrate unit cells coincide and hence these directions remain reflection axes for the superlattice. As a result, $\theta=0$ requires $U_{G'}=0$. \cite{wallbank_prb_2013_root3}

The reciprocal lattice vectors $\{\vect{\beta}_{m}\}$ correspond to the real space periodicity set by the lattice vectors of the substrate and the Kekul\'{e} lattice of graphene, depicted by vector $\vect{A}$ in Fig.~\ref{fig:figure1}(a). However, we demonstrate in the Appendix that a shorter periodicity of the local atomic arrangement surrounding a carbon atom, indicated in Fig.~\ref{fig:figure1}(a) by vector $\vect{\tilde{A}}$, always exists. This shorter periodicity corresponds to primitive reciprocal lattice vectors $\{\vect{b}_{m}\} = \sqrt{3} \hat{R}_{\frac{\pi}{2}} \vect{\beta}_{m}$, shown in Fig. 1(b), which define a larger superlattice Brillouin zone (sBZ), shown in black in Fig.~\ref{fig:figure1}(b), than the vectors $\{\vect{\beta}_{m}\}$. \cite{wallbank_prb_2013_root3} Existence of the two periodicities gives rise to the particular combination of functions $f_{1}(\vect{\check{\beta}})$ and $f_{2}(\vect{\check{\beta}})$ in Eq.~\eqref{HamilAndPert} which always leads to an exact cancellation of half of the terms in the sums over vectors $\{\vect{\beta}_{m}\}$. As a result, a state $\ket{+,n,\vect{p}}$ with momentum $\vect{p}$ in the vicinity of the valley $\vect{K}_{+}$ and with $n$ indexing one of the four BLG bands, is directly coupled by the moir\'{e} perturbation to three states $\ket{-,n',\vect{p}+\vect{\beta}_{m}}$, $m=1,3,5,$ [shown in green in Fig.~\ref{fig:figure1}(b)] in the vicinity of the valley $\vect{K}_{-}$. Equivalently, a state $\ket{-,n,\vect{p}}$ is coupled to the states $\ket{+,n',\vect{p}+\vect{\beta}_{m}}$, $m=0,2,4$ [shown in red in Fig.~\ref{fig:figure1}(b)]. Hence, in the reduced zone scheme, the centre of the $\vect{K}_{-}$ valley is folded onto momentum $\vect{\beta}_{0}$ in the vicinity of the valley $\vect{K}_{+}$. In this work, we discuss the miniband spectrum within the sBZ set by the vectors $\{\vect{b}_{m}\}$, denoting its high-symmetry points as shown in Fig.~\ref{fig:figure1}(b). In order to treat both valleys on an equal footing, we choose the position of the sBZ which reflects best the symmetry of the lattice: the centre of the valley $\vect{K}_{+}$ is at the point $\vect{\kappa}$ and $\vect{K}_{-}$ is mapped onto $\vect{\kappa}'$.\cite{wallbank_prb_2013_root3} The result of such folding of the unperturbed BLG spectrum is shown in Fig.~\ref{fig:figure1}(c), where we depicted bands from the $\vect{K}_{+}$ ($\vect{K}_{-}$) valley in magenta (yellow). We calculate the miniband spectra like the one shown in Fig.~\ref{fig:figure1}(c) by numerical diagonalization of the Heisenberg matrix built of 31 points coupled by the moir\'{e} perturbation, including the initial point $\ket{+,n,\vect{p}}$ (mixing of the valleys ensures that the miniband structure calculated for the choice of the initial point $\ket{-,n,\vect{k}}$ is identical). We take into account states in all four of the BLG $\pi$-bands and to set the geometry, we choose ${\mathrm{In}}_{2}{\mathrm{Te}}_{2}$, a semiconductor with a band gap of $\sim 2$eV, \cite{zolyomi_prb_2014_inte} as the intended substrate (a list and a short discussion of other potential substrates can be found in Ref.~\onlinecite{wallbank_prb_2013_root3}). This sets the lattice mismatch $\delta=-0.007$ \cite{zolyomi_prb_2014_inte, giovannetti_prb_2015} and the characteristic energy of the moir\'{e} $\sqrt{3}v\beta=vb=0.134$eV. 

\section{Effective Hamiltonians at high-symmetry points}

In this section, we assume that $\tfrac{vb}{\gamma_{1}}<1$ (which requires small lattice mismatch $\delta$) and, because we are interested in the reconstruction of the electronic spectrum at the boundary of the first and second miniband, we ignore the high-energy split bands. We also assume that $\theta=0$ and hence $U_{G'}=0$.\cite{comment_angle_dependence} We write the unperturbed plane wave state $\ket{\xi,s,\vect{p}}$ in the conduction ($s=1$) or valence ($s=-1$) band, with momentum $\vect{p}=(p_{x},p_{y})\neq 0$ in the vicinity of the valley $\vect{K}_{\xi}$,
\begin{align*}\begin{split}
&\ket{+,s,\vect{p}} = \frac{1}{\sqrt{C_{p}}} \begin{pmatrix} \ket{+}\otimes\vect{u}_{+} \\ \ket{+}\otimes\vect{v}_{+} \end{pmatrix} e^{i\vect{p}\cdot\vect{r}}, \\
&\ket{-,s,\vect{p}} = \frac{1}{\sqrt{C_{p}}} \left(\begin{matrix} \ket{-}\otimes\sigma_{x}\vect{u}_{-} \\ \ket{-}\otimes\sigma_{x}\vect{v}_{-} \end{matrix}\right) \! e^{i\vect{p}\cdot\vect{r}}, \\
& \vect{u}_{\xi} = \begin{pmatrix} 1 \\ \dfrac{1}{v p} (\epsilon_{p,s}^{0} - \frac{u}{2}) e^{i\xi\phi} \end{pmatrix},\\
& \vect{v}_{\xi} = \frac{(\epsilon_{p,s}^{0} - \frac{u}{2})^{2} - v^{2} p^{2}}{\gamma_{1}}\begin{pmatrix} \dfrac{1}{v p} e^{i\xi\phi} \\ \dfrac{1}{(\epsilon_{p,s}^{0} + \frac{u}{2})} e^{i\xi 2\phi} \end{pmatrix},\\
& \epsilon_{p,s}^{0} =s \sqrt{\frac{ \gamma_{1}^{2}}{2} + \frac{u^{2}}{4} +v^{2} p^{2} - \sqrt{\frac{\gamma_{1}^{4}}{4} + v^{2} p^{2} (\gamma_{1}^{2} + u^{2})}},
\end{split}\end{align*}
where $\tan\phi=p_{y}/p_{x}$, $p=\sqrt{p_{x}^{2}+p_{y}^{2}}$, $\ket{+}=(1,0)^{T}$, $\ket{-}=(0,1)^{T}$ and $C_{p}$ is a normalisation constant. 

At the $\vect{\mu}$ point, zone folding brings together two degenerate states, $\ket{+,s,\tfrac{\vect{\beta}_{0}}{2}}$ and $\ket{-,s,-\tfrac{\vect{\beta}_{0}}{2}}$. Applying degenerate perturbation theory to these two states leads to a $2\times 2$ matrix,
\begin{align*}\begin{split} 
& \op{H}_{\mu}=\begin{pmatrix}
\epsilon_{\beta/2,s}^{0}  & \Delta_{\mu,s} \\
\Delta_{\mu,s}  & \epsilon_{\beta/2,s}^{0} \\
\end{pmatrix}, \\
& \Delta_{\mu,s}=  -\frac{2vb}{\sqrt{3}C_{\mu}} \bigg\{ 4\sqrt{3} U_{E'} \frac{\left( \epsilon_{\beta/2,s}^{0} - \frac{u}{2} \right)}{vb} \\ & - U_{G} \bigg[ 12 \frac{\left( \epsilon_{\beta/2,s}^{0} - \frac{u}{2} \right)^{2}}{v^{2}b^{2}} + 1 \bigg]  \bigg\},
\end{split}\end{align*}
and yields the perturbed energies
\begin{align}\label{eqn:mu}
\epsilon_{\mu,s}^{\pm}=\epsilon_{\beta/2,s}^{0}\pm|\Delta_{\mu},s|. 
\end{align}

At the $\vect{\gamma}$ point, the six following degenerate states are mixed together by the perturbation: $\ket{+,s,\vect{\beta}_{5}}$,  $\ket{-,s,\vect{\beta}_{0}}$, $\ket{+,s,\vect{\beta}_{1}}$, $\ket{-,s,\vect{\beta}_{2}}$, $\ket{+,s,\vect{\beta}_{3}}$ and $\ket{-,s,\vect{\beta}_{4}}$. Only the neighbours in the effective ring of six points are directly coupled and the couplings are related by a phase, yielding a $6\times 6$ matrix,
\begin{align*}
& \op{H}_{\gamma}\!=\!\begin{pmatrix}
\epsilon_{\beta,s}^{0} & \Delta_{\gamma,s}w^{*} & 0 & 0 & 0 & -\Delta_{\gamma,s} \\
\Delta_{\gamma,s}w & \epsilon_{\beta,s}^{0} &  \Delta_{\gamma,s}w^{*} & 0 & 0 & 0 \\
0 & \Delta_{\gamma,s}w & \epsilon_{\beta,s}^{0} & -\Delta_{\gamma,s} & 0 & 0 \\
0 & 0 & -\Delta_{\gamma,s} & \epsilon_{\beta,s}^{0} & \Delta_{\gamma,s}w & 0 \\
0 & 0 & 0 & \Delta_{\gamma,s}w^{*} & \epsilon_{\beta,s}^{0} & \Delta_{\gamma,s}w \\
-\Delta_{\gamma,s} & 0 & 0 & 0 & \Delta_{\gamma,s}w^{*} & \epsilon_{\beta,s}^{0} \\
\end{pmatrix}\!, \\
& \Delta_{\gamma,s} \!=\! \frac{2vb}{\ \!\sqrt{3}C_{\gamma}} \bigg\{ \sqrt{3} U_{E'} \frac{\left( \epsilon_{\beta,s}^{0} \!-\! \frac{u}{2} \right)}{vb} \!-\! U_{G} \bigg[ 3 \frac{\left( \epsilon_{\beta,s}^{0} \!-\! \frac{u}{2} \right)^{2}}{v^{2}b^{2}} \!+\! 1 \bigg] \bigg\},\nonumber\\
& w = \exp\left(i\frac{\pi}{3}\right).\nonumber
\end{align*}
As a result of the perturbation, the six levels split into two degenerate pairs of levels and two nondegenerate states,
\begin{align}\label{eqn:gamma}
\epsilon_{\gamma,s}^{\pm,\mathrm{deg}}=\epsilon_{\beta,s}^{0}\pm|\Delta_{\gamma,s}|, \,\,\,\,\,\,\,\, \epsilon_{\gamma,s}^{\pm}=\epsilon_{\beta,s}^{0}\pm2|\Delta_{\gamma,s}|.
\end{align}

Finally, at the $\vect{\kappa}$/$\vect{\kappa}'$ point, two (degenerate for $u=0$) states at the centre of the valley $\xi$,
\begin{align*}\begin{split}
&\ket{\xi,\vect{0}}_{i} =\begin{pmatrix} \ket{\xi}\otimes\begin{pmatrix} \delta_{i,1}\delta_{\xi,1} \\ \delta_{i,1}\delta_{\xi,-1} \end{pmatrix} \\ \ket{\xi}\otimes\begin{pmatrix} \delta_{i,2}\delta_{\xi,-1} \\ \delta_{i,2}\delta_{\xi,1} \end{pmatrix} \end{pmatrix},\,\,\,\,\,\,i=1,2,
\end{split}\end{align*}
where $\delta_{ij}$ is the Kronecker delta, are each coupled to six points $\ket{-\xi,\pm 1,\vect{\beta}_{j}}$, where $j$ is odd if $\xi=1$ and even for $\xi=-1$. However, the moir\'{e} perturbation acts only on the bottom graphene layer, see Eq.~\eqref{HamilAndPert}, while the state $\ket{\xi,\vect{0}}_{2}$ is located exclusively on the top layer and, hence, is effectively uncoupled from the other states, its energy, $\epsilon=-\tfrac{u}{2}$, not affected by the moir\'{e} perturbation. The remaining 7 points lead to the matrix,
\begin{align*}
& \op{H}_{\kappa}\!=\!\begin{pmatrix}
\frac{u}{2} & \op{T}_{1} & \op{T}_{2} \\
\op{T}_{1}^{\dagger} & \op{H}_{\kappa,1}^{0} & 0 \\
\op{T}_{2}^{\dagger} & 0 & \op{H}_{\kappa,-1}^{0} \\
\end{pmatrix},\,\,\, 
\op{H}_{\kappa,s}^{0} \! = \! \begin{pmatrix}
\epsilon_{\beta,s}^{0} & 0 & 0 \\
0 & \epsilon_{\beta,s}^{0} & 0 \\
0 & 0 & \epsilon_{\beta,s}^{0} \\
\end{pmatrix}, \\
& \op{T}_{1} = \begin{pmatrix}
 \Delta_{\kappa,1}w & -\Delta_{\kappa,1} & \Delta_{\kappa,1}w^{*} \\
\end{pmatrix}, \\
& \op{T}_{2} = \begin{pmatrix}
 \Delta_{\kappa,-1}w & -\Delta_{\kappa,-1} & \Delta_{\kappa,-1}w^{*} \\
\end{pmatrix}, \\
&\Delta_{\kappa,s} = \frac{2vb}{\sqrt{3 C_{\beta}}} \bigg\{ \sqrt{3} U_{E'} \frac{\left(\epsilon_{\beta,s}^{0} - \frac{u}{2} \right)}{vb} - U_{G} \ \bigg\}.\\
\end{align*}
For both $u\ll\epsilon_{\beta,s}^{0}$ and $|\Delta_{\kappa,s}|\ll\epsilon_{\beta,s}^{0}$, we can use the Schrieffer-Wolff transformation \cite{schrieffer_physrev_1966} to project the Hamiltonian above onto the low-energy state $\ket{\xi,\vect{0}}_{1}$. As a result, we obtain the shift of its energy from $\epsilon=\tfrac{u}{2}$ due to the perturbation, 
\begin{align}\label{eqn:kappa}
\epsilon_{\kappa}\approx\frac{u}{2} +8\sqrt{3}vbU_{E'}U_{G}.
\end{align}

\section{Miniband spectrum}

We are interested in determining the conditions for which the first and second miniband on the conduction/valence-band side are separated by a gap. In what follows, we discuss the valence-band side; the respective conditions for the conduction-band side can be obtained by taking advantage of the symmetry of the miniband spectrum,
\begin{align}
\epsilon^{U_{E'},U_{G},U_{G'},u}_{\vect{k}} = -\epsilon^{-U_{E'},U_{G},U_{G'},-u}_{\vect{k}}.
\end{align}

In the presence of a weak perturbation, in order to confirm the presence of a band gap between the first and second miniband, it is enough to analyse energy states at the high-symmetry points. Ignoring the Mexican-hat features created at the valence band edge by nonzero interlayer asymmetry $u$, \cite{mccann_prb_2006_extpot} the highest (closest to the neutrality point) point in the first miniband is the centre of the valley at $\vect{\kappa}/\vect{\kappa}'$ at energy $\epsilon=\mathrm{min}(\epsilon_{\kappa},-\tfrac{u}{2})$. Because the unperturbed BLG dispersion has circular symmetry and electron energy in the valence band decreases away from the centre of the valley, the point with the lowest energy in the first miniband is that furthest away from the centre of the valley, the $\vect{\gamma}$ point, with energy $\epsilon_{\gamma,-1}^{+}$, Eq.~\eqref{eqn:gamma}. In turn, the point with the highest energy in the second miniband is the $\vect{\mu}$ point which lies in the middle of the shortest line segment connecting two valleys, with energy $\epsilon_{\mu,-1}^{-}$, Eq.~\eqref{eqn:mu}.

\begin{figure}
\subfloat{  \includegraphics[clip,width=\columnwidth]{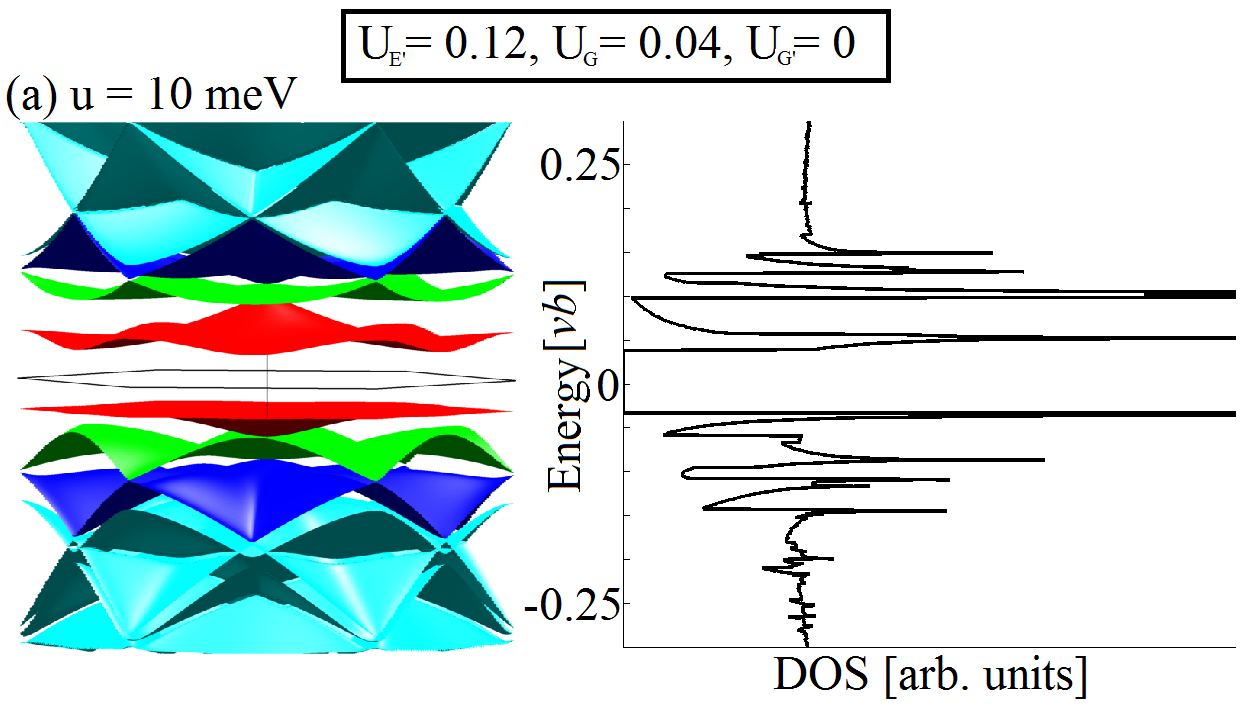} } \\ \hspace*{-0.2em}
\subfloat{  \includegraphics[clip,width=\columnwidth]{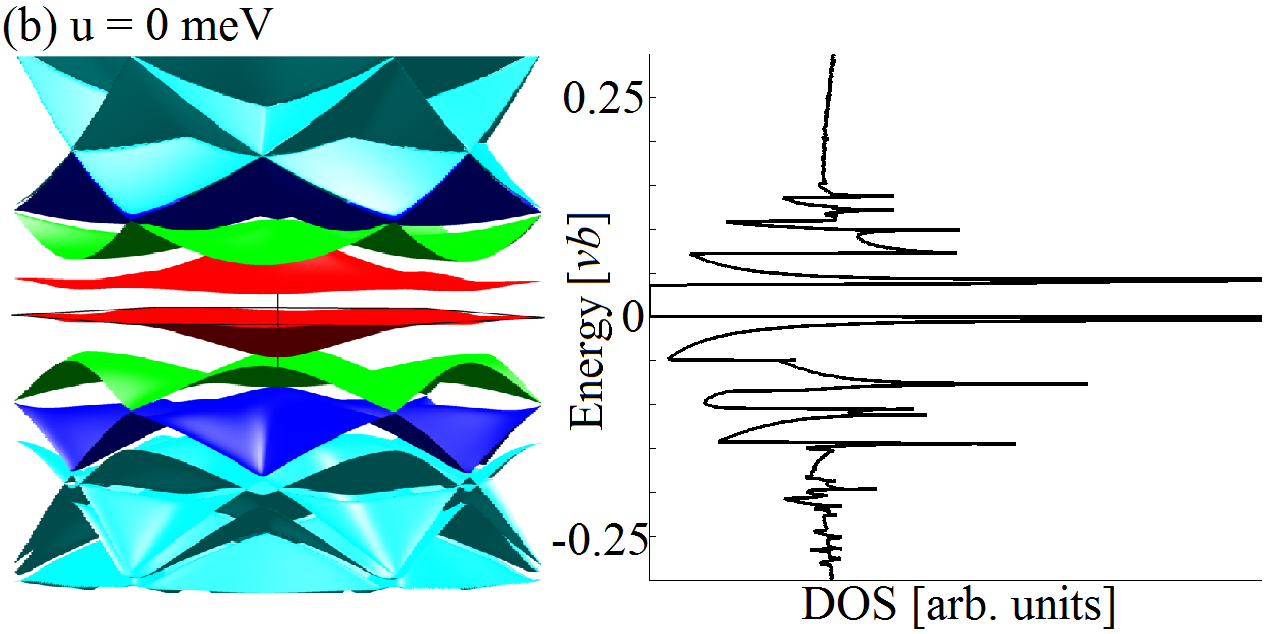} } \\ \hspace*{-0.2em}
\subfloat{  \includegraphics[clip,width=\columnwidth]{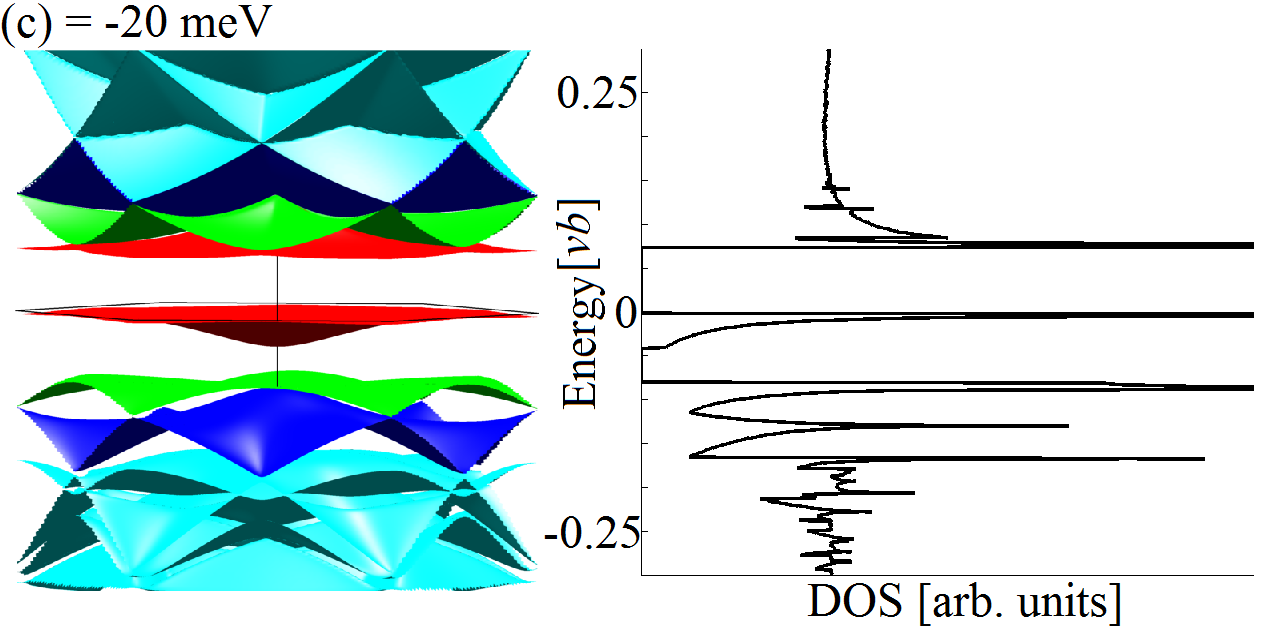} } \\ 
\caption{(a)–(c) Moir\'{e} miniband spectra and density of states (DoS) corresponding to characteristic behaviors of the miniband spectrum, as discussed in the text.}
\label{fig:figure2}
\end{figure}

In contrast to monolayer graphene, \cite{wallbank_prb_2013_root3} in BLG the energies of the extremal points of a miniband can be modified by tuning the interlayer asymmetry parameter, $u$, through application of an external electric field perpendicular to the graphene layers. Nonzero $u$ is known to open a gap at the neutrality point. Here, depending on the sign and magnitude of $u$, one could open and close an additional band gap in the band structure, between the first and second miniband. An example of such tuning of the miniband spectrum is shown in Fig.~\ref{fig:figure2}, where we show the miniband spectra for $U_{E'}=0.12$, $U_{G}=0.04$ and $U_{G'}=0$ and three different values of $u$, with the additional band gap visible in the spectrum in (c). For $u=10$ meV, Fig.~\ref{fig:figure2}(a), the miniband spectrum contains a single band gap at the neutrality point, $E_{g}\approx u$. As the interlayer asymmetry is decreased, the band gap decreases and for $u=0$, Fig.~\ref{fig:figure2}(b), this gap is due only to the moir\'{e} perturbation, $E_{g}\approx 8\sqrt{3}vbU_{E'}U_{G}$. Reversal of the sign of $u$ leads to a closure of the band gap when the effect of the electric field cancels that of the moir\'{e} perturbation. Further increase of the magnitude of $u$, Fig.~\ref{fig:figure2}(c), again opens a band gap at the neutrality point. However, it also opens a band gap, $E_{g}'\approx |\Delta_{\mu,-1}| + 2 |\Delta_{\gamma,-1}|$, between the first and second miniband in the valence band. 

The presence of band gaps in the miniband spectrum can be further confirmed through investigation of the density of states (DoS), shown to the right of each of the miniband spectra in Fig.~\ref{fig:figure2}. Accordingly, in Fig.~\ref{fig:figure2}(c), the DoS vanishes both in the vicinity of the neutrality point as well as below the first Van Hove singularity in the valence band. 

\begin{figure}
\subfloat{  \includegraphics[clip,width=0.8\columnwidth]{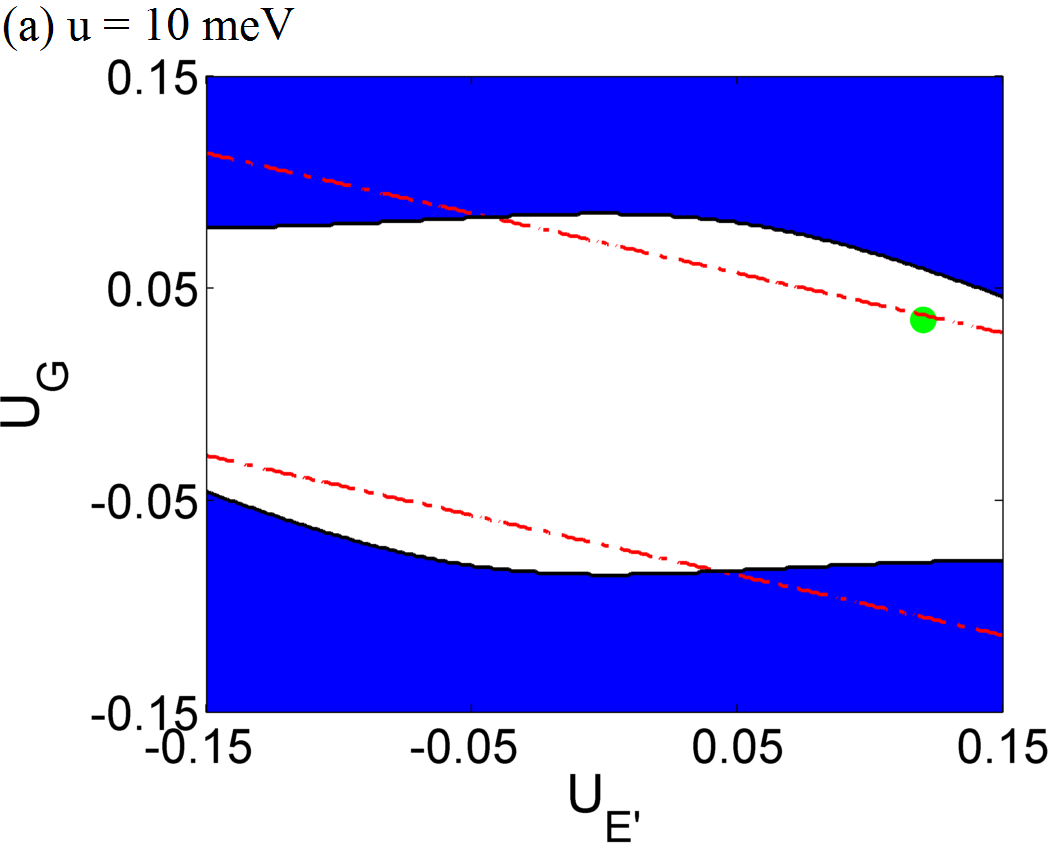} } \\ \hspace*{-0.2em}
\subfloat{  \includegraphics[clip,width=0.8\columnwidth]{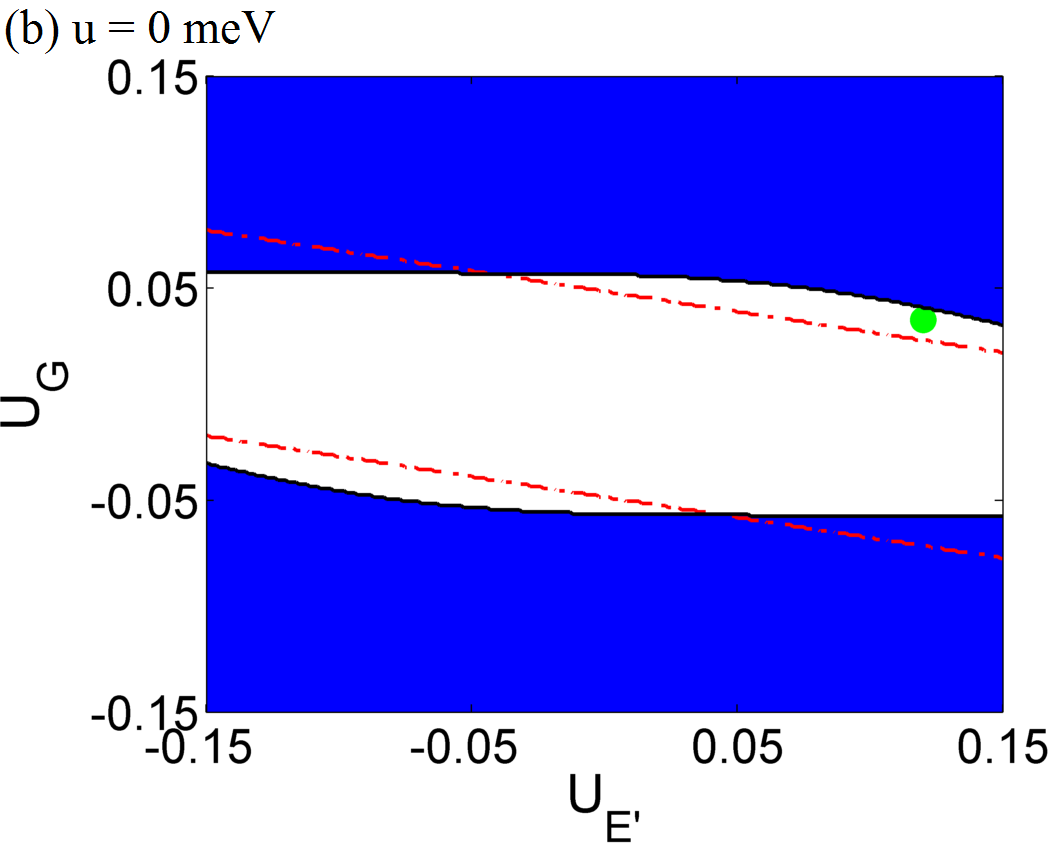} } \\ \hspace*{-0.2em}
\subfloat{  \includegraphics[clip,width=0.8\columnwidth]{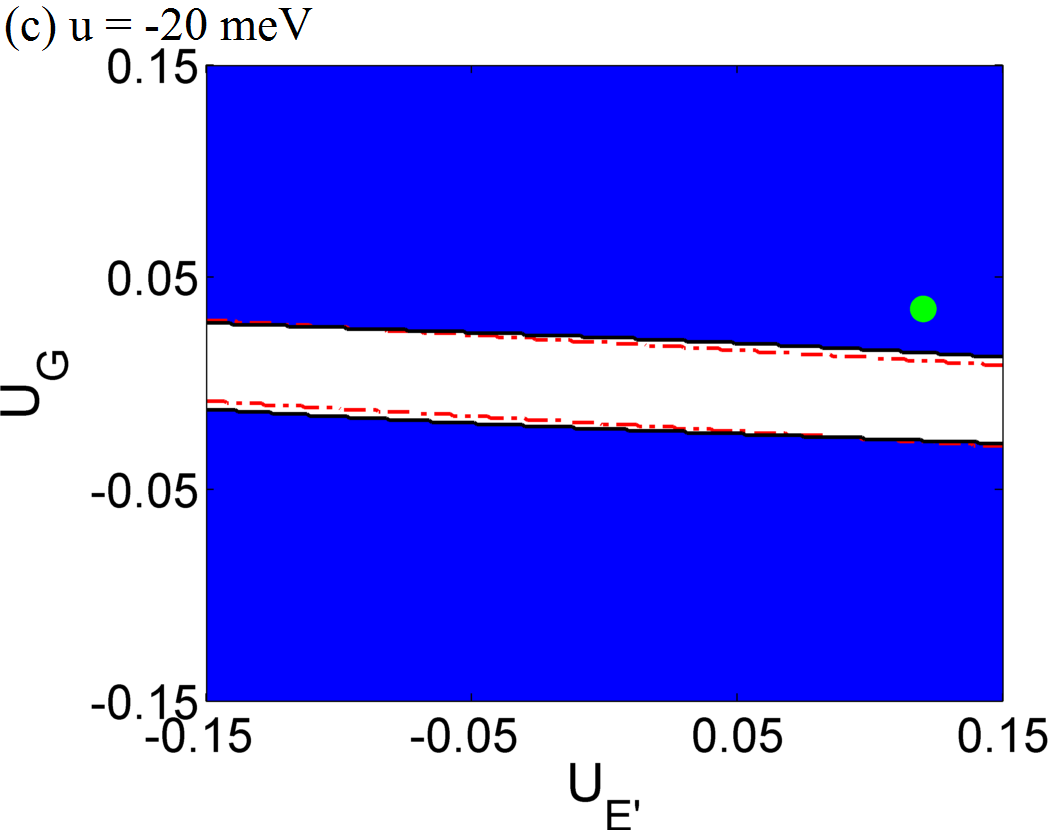} } \\
\caption{Regimes of miniband spectra in the $(U_{E'},U_{G})$ parameter space for the heterostructure of BLG and $\mathrm{In}_{2}\mathrm{Te}_{2}$ ($\delta=-0.007$) and (a) $u=10$, (b) $u=0$ and (c) $u=-20$ meV. For the region in blue, a global gap separates the first and seconf miniband on the valence side. The white region corresponds to the overlapping first and second minibands. The dashed red line represents the boundary between the blue and white regions as obtained using first order perturbation theory analysis of the high-symmetry points in the sBZ. The green dot represents point in the $(U_{E'},U_{G})$ space for which miniband spectra are shown in Fig.~\ref{fig:figure2}.
Fig. 2.}
\label{fig:figure3}
\end{figure}

In Fig.~\ref{fig:figure3}, we show the results of our study of the miniband spectrum for a generic moir\'{e} perturbation described by parameters $U_{E'}$ and $U_{G}$ for various $u$. The blue-coloured region (I) shows the range of parameters $U_{E'}$ and $U_{G}$ for which a band gap between the first and second miniband in the valence band exists. The region (II) in white corresponds to the case of overlapping minibands. While the diagram has been produced by inspecting numerically calculated miniband structures, the red dashed line indicates the boundary between the regions (I) and (II) as predicted by the analytical considerations presented in the previous section and inspection of the energy states at the high-symmetry points. The miniband spectra shown in Fig.~\ref{fig:figure2} correspond to the moir\'{e} parameter set indicated by the green dot in the diagrams in Fig.~\ref{fig:figure3}(a)-(c). 

In general, greater magnitudes of the perturbation parameters promote appearance of the band gap between the first and second miniband for smaller $u$. However, the effect of the parameter $U_{G}$, describing sublattice-conserving part of the moir\'{e} perturbation, is more significant than that of the parameter $U_{E'}$, characterising the sublattice-exchange part of the perturbation. Also, the threshold value of the interlayer asymmetry $u$ necessary to open the band gap between the first and second miniband for a set moir\'{e} perturbation is different for different signs of $u$.  

\section{Summary}

We have discussed the generic miniband structure of the van der Waals heterostructure of bilayer graphene and a semiconducting substrate almost commensurate with the tripled unit cell of graphene. We showed that the combination of an external electric field normal to the graphene layers, which modifies the band structure in the vicinity of the neutrality point, and the miniband formation due to the substrate lead to novel allow new degree of tunability of graphene band structure: the miniband structure can be tuned from gapless to that displaying two band gaps, one at the neutrality point and one between the first and second miniband on the conduction/valence-band side, hence isolating a single miniband from the rest of the spectrum. For the case of the lattice mismatch $\delta=-0.007$, corresponding to the choice of $\mathrm{In}_{2}\mathrm{Te}_{2}$ as the substrate, and mismatch angle $\theta=0$, such an isolated miniband could be realised by using relatively weak, experimentally accessible electric fields. For this particular substrate, the isolated valence miniband would have a width of $t\sim\tfrac{(vb)^{4}}{9\gamma_{1}^{2}|u|}$ and accommodate carrier density $n_{0}=\frac{\sqrt{3} b^{2}}{4 \pi^{2}}\approx 1.9 \times 10^{11}$cm$^{-2}$.

\begin{acknowledgments}

The authors would like to thank J.~R.~Wallbank for his comments on the manuscript. This work has been supported by the EPSRC through the grant EP/L013010/1 and the University of Bath Doctoral Training Account, EP/M50645X/1.

\end{acknowledgments}

\vspace{10pt}

\appendix*

\section{Superlattice periodicities for graphene on incommensurate $\sqrt{3}\times\sqrt{3}$ substrate}

In this appendix, we demonstrate that presence of the moir\'{e} superlattice set by the graphene Kekul\'{e} lattice and the substrate implies that a superlattice with a shorter period, generated by the substrate and the graphene lattice, also exists.
 
The substrate is almost commensurate with the Kekul\'{e} lattice of graphene and as a result, a new long wavelength periodicity forms at the interface between the two crystals. A continuum model can be constructed if the structure is close to a commensurate geometry \cite{wallbank_prb_2013, lopes_dos_santos_prl_2007} and here we assume that the unit vector $\vect{A}$ of the superlattice can be expressed in terms of the lattice vectors of the substrate and graphene Kekul\'{e} lattice so that
\begin{align}\begin{split} \label{BetaReq}
\vect{A} & = m\left( 2 \vect{a}_{1} + \vect{a}_{2} \right) + n \left( \vect{a}_{1} + 2 \vect{a}_{2} \right) \\
& = \sqrt{1 + \delta} \op{R}_{\theta} \!\left( p \tilde{\vect{a}}_{1} + q \tilde{\vect{a}}_{2} \right),
\end{split}\end{align}
where $\vect{a}_{1}=a(\tfrac{1}{2},\tfrac{\sqrt{3}}{2})$ and $\vect{a}_{2}=a(\tfrac{1}{2},-\tfrac{\sqrt{3}}{2})$ are the unit vectors of graphene, $\tilde{\vect{a}}_{1}=a(\tfrac{3}{2},\tfrac{\sqrt{3}}{2})$ and $\tilde{\vect{a}}_{2}=a(\tfrac{3}{2},\tfrac{-\sqrt{3}}{2})$ are the unit vectors of the Kekul\'{e} lattice of graphene and $m$, $n$, $p$ and $q$ are integers. This periodicity gives rise to the primitive reciprocal vectors $\{\vect{\beta}_{m}\}$.

The above condition can be cast in the matrix form 
\begin{gather*}
\begin{pmatrix}
m \\ n
\end{pmatrix} \! = \! \sqrt{1+\delta}\! \begin{pmatrix}
\cos\theta + \frac{1}{\sqrt{3}} \sin\theta & \frac{2}{\sqrt{3}}\sin\theta \\
-\frac{2}{\sqrt{3}} \sin\theta & \cos\theta - \frac{1}{\sqrt{3}}\sin\theta \\
\end{pmatrix}•\! \begin{pmatrix}
p \\ q
\end{pmatrix},
\end{gather*}
mapping one pair of integers to another. The necessary and sufficient condition for that \cite{shallcross_prb_2010} is that the elements of the matrix above assume rational values,
\begin{equation} \label{eqn:rationality}
\sqrt{1+\delta}\frac{1}{\sqrt{3}}\sin\theta = \frac{i_{1}}{i_{3}}, \sqrt{1+\delta}\cos\theta = \frac{i_{2}}{i_{3}}.
\end{equation}

Let us now investigate a vector $\tilde{\vect{A}}$, equivalent to the vector $\vect{A}$ from Eq.~\eqref{BetaReq} rotated by $\tfrac{\pi}{6}$ and shorter by $\tfrac{1}{\sqrt{3}}$. We have for the left-hand side
\begin{align*}
\tilde{\vect{A}} = & \frac{1}{\sqrt{3}} \op{R}_{\pi/6} \left\{ m \left( 2 \vect{a}_{1} + \vect{a}_{2} \right) + n \left( \vect{a}_{1} + 2 \vect{a}_{2} \right) \right\} \\
= & \left( m + n \right) \vect{a}_{1} + n \vect{a}_{2}.
\end{align*}
For the vector above to describe a periodic structure formed by the lattice of the substrate and the lattice of graphene, we require similarly to Eq.~\eqref{BetaReq} that 
\begin{equation*}
\left( m + n \right) \vect{a}_{1} + n \vect{a}_{2} = \sqrt{1 + \delta} \op{R}_{\theta} \left( P \tilde{\vect{a}}_{1} + Q \tilde{\vect{a}}_{2} \right),
\end{equation*}
where $P$ and $Q$ are integers. This can be also written as
\begin{gather*}
\begin{pmatrix}
m \\ n
\end{pmatrix} \!=\! \sqrt{1+\delta}\! \begin{pmatrix}
\cos\theta \!+\! \sqrt{3} \sin\theta & -\cos\theta \!+\! \sqrt{3} \sin\theta \\
\cos\theta \!-\! \sqrt{3} \sin\theta & 2\cos\theta \\
\end{pmatrix} \! \begin{pmatrix}
P \\ Q
\end{pmatrix}.
\end{gather*}
Notice that the entries in the matrix above have to be rational because, from Eq.~\eqref{eqn:rationality},
\begin{align}
\sqrt{1+\delta}\sqrt{3}\sin\theta=3\frac{i_{1}}{i_{3}},\,\,\,\,\sqrt{1+\delta}\cos\theta=\frac{i_{2}}{i_{3}}.
\end{align}
Hence, if the substrate and Kekul\'{e} lattices form a superlattice with a unit vector $\vect{A}$, then the substrate and graphene lattices form a superlattice with a unit vector $\tilde{\vect{A}}=\tfrac{1}{\sqrt{3}}\op{R}_{\pi/6}\vect{A}$, as shown in Fig.~\ref{fig:figure1}(a). The primitive reciprocal lattice vectors corresponding to the latter, shorter periodicity are rotated by $\tfrac{\pi}{6}$ with respect to $\{\vect{\beta}_{m}\}$ and $\sqrt{3}$ times longer what is equivalent to the definition of the vectors $\{\vect{b}_{m}\}$ in the main text.


\begin{thebibliography}{99}
\bibitem{comment_angle_dependence} Note that, in the case of the graphene/hBN heterostructure, the effect of the moir\'{e} on graphene electrons is most pronounced for small misalignment angles $\theta$ (Ref.~\onlinecite{wallbank_adp_2015}) and the same should be the case for the heterostructure considered here.
\end{thebibliography}
\end{document}